\documentclass[journal,twocolumn]{IEEEtran}
\usepackage{}
\usepackage{graphicx}
\usepackage{epstopdf}
\usepackage{amssymb}
\usepackage{amsfonts}
\usepackage{amsmath}
\usepackage{algorithm}
\usepackage{algorithmic}
\usepackage{subeqnarray}
\usepackage{cases}

\usepackage{subfigure,amsmath,amssymb,cite}
\usepackage{float}
\ifCLASSINFOpdf
\else
\fi
\begin{document}

\title{Performance Analysis of Discrete-Phase-Shifter IRS-aided Amplify-and-Forward Relay Network}

\author{Rongen Dong,~Zhongyi Xie,~Feng Shu,~Mengxing Huang,~and Jiangzhou Wang,\emph{ Fellow, IEEE}

\thanks{This work was supported in part by the National Natural Science Foundation of China (Nos.U22A2002, and 62071234), the Hainan Province Science and Technology Special Fund (ZDKJ2021022), the Scientific Research Fund Project of Hainan University under Grant KYQD(ZR)-21008, and the Collaborative Innovation Center of Information Technology, Hainan University (XTCX2022XXC07).}
\thanks{Rongen Dong, Zhongyi Xie, and Mengxing Huang are with the School of Information and Communication Engineering, Hainan University, Haikou, 570228, China.}
\thanks{Feng Shu is with the School of Information and Communication Engineering and Collaborative Innovation Center of Information Technology, Hainan University, Haikou 570228, China, and also with the School of Electronic and Optical Engineering, Nanjing University of Science and Technology, Nanjing 210094, China (Email: shufeng0101@163.com).}
\thanks{Jiangzhou Wang is with the School of Engineering, University of Kent, Canterbury CT2 7NT, U.K. (Email: {j.z.wang}@kent.ac.uk).}



}

\maketitle

\begin{abstract}
As a new technology to reconfigure wireless communication environment by signal reflection controlled by software, intelligent reflecting surface (IRS) has attracted lots of attention in recent years. Compared with conventional relay system, the relay system aided by IRS can effectively save the cost and energy consumption, and significantly enhance the system performance. However, the phase quantization error generated by IRS with discrete phase shifter may degrade the receiving performance of the receiver. To analyze the performance loss arising from IRS phase quantization error, in accordance with the law of large numbers and Rayleigh distribution, the closed-form expressions for the signal-to-noise ratio (SNR) performance loss and achievable rate of the double IRS-aided amplify-and-forward (AF) relay network, which are associated with the number of phase shifter quantization bits, are derived in the Rayleigh channels. In addition, their approximate performance loss closed-form expressions are also derived based on the Taylor series expansion. Simulation results show that the performance losses of SNR and achievable rate decrease with the number of quantization bits increases gradually, and increase with the number $k$ of IRS phase shift elements. The SNR and achievable rate performance losses of the system are smaller than 0.06dB and 0.03bits/s/Hz when $k$ is equal to 4 and 3, respectively.
\end{abstract}


\section{INTRODUCTION}
With the advent of the sixth-generation (6G) era, ubiquitous wireless network has become a reality, and relay has been widely adopted as an effective way to boost the quality of wireless communication. When long-distance end-to-end communication is performed, the relay can be used as transmission nodes to assist communication as a means to enhance the transmission performance of the system \cite{Cover1979Capacity}. The relay selection technology can meet the demand for diversity gain and reduce the degree of fading of the user's received signal\cite{Ding2010Diversity}. For example, the signal space diversity-dual-hop cooperative system with an energy harvesting and decode-and-forward (DF) relay was considered in \cite{Ammar2021Signal}, and the exact closed-form expression of the outage probability was derived. The traversal capacity of multi-hop wireless communication network with amplify-and-forward (AF) relay was investigated in \cite{Farhadi2009On}, and two upper bounds on the traversal capacity of AF multi-hop communication system was obtained.

However, deploying a large number of active devices in the wireless communication system would result in a large amount of energy consumption. Being an energy-efficient and effective solution, intelligent reflecting surface (IRS), as a flat surface made up of a large number of low-cost passive reflective elements, each of which can independently tune the phase and/oramplitude of the incident signal\cite{Shu2021Enhanced}. As a potential key technology for future 6G wireless network, IRS brings a new communication network paradigm by creating an intelligent and controllable wireless environment. IRS can improve the wireless transmission environment and effectively tackle the problems of signal fading caused by the large-scale application of millimeter wave communication technology \cite{Di2020Smart}.

An effective combination of relay and IRS has been proven to be an effective means of enhancing the spectral and energy efficiencies, as well as rate performance\cite{ Yildirim2021Hybrid}.
To investigate the advantages of combining conventional DF relay system with IRS, a new hybrid relay and IRS-assisted system was proposed in \cite{Abdullah2020A}. Compared to\cite{Abdullah2020A}, \cite{Wang2022Beamforming} increased the number of antennas at the relay, and maximized the received power by alternately optimizing the precoder at the relay and the phase shift matrix at IRS, in this IRS-assisted multi-antenna relay model, a null-space projection plus maximum ratio combining method was proposed to enhance the SNR at the relay.
The performance of IRS-assisted AF relay system was investigated in \cite{Galappaththige2021On}, to probabilistically characterize optimal SNR, a tightly approximated cumulative distribution function was derived. When there is an untrusted relay, the AF relay may eavesdrop while helping to forward message. To maximize the secrecy rate of the IRS-aided system when there was an untrusted relay, an alternating iteration scheme was proposed to jointly design the active and passive beamforming\cite{Liu2023IRS}.

However, all the above works were performed based on IRS with continuous phase shifters, i.e., no phase quantization error. In practice, since there are the hardware constraints, IRS can only be operated with discrete phase shifters\cite{You2020Channel}. Similar to the performance loss caused by the use of finite quantization bit radio frequency phase shifters and hybrid phase shifters in directional modulation network\cite{Li2019Performance, Dong2022Performance}, the IRS with discrete phase shifters yields phase quantization error that may degrade the performance of the receiver\cite{Wu2020Beamforming, Dong2022Performance2}.
To analyze the performance loss arising from IRS phase quantization error in double IRS-aided AF relay network and to provide a reference for choosing IRS with proper number of quantization bits in practice, in this paper, we analyze both the performance losses and approximate performance losses of the signal-to-noise ratio (SNR) and achievable rate (AR) in the Rayleigh channels. The main contributions of this paper are summarized as follows:
\begin{enumerate}
\item Aiming to make an analysis of the effect of discrete phase shifters IRS on system performance and boost the rate of conventional relay network, a double IRS-assisted AF relay system is considered, where the number of phase shift elements may be different for IRS-1 and IRS-2. It is assumed that all channels are the Rayleigh channels. In accordance with the weak law of large numbers, Euler formula, and Rayleigh distribution, the closed-form expression of the SNR performance loss, as the functions of the number of quantization bits of the phase shifter, is derived. Moreover, the approximate performance loss closed-form expression with phase quantization error is also derived by utilizing the first-order Taylor expansion.
\item  In the same manner, the closed-form expressions for the achievable rate with no performance loss, with performance loss, and with approximate performance loss scenarios are also derived. From the simulation results, it can be found that the SNR performance loss and the achievable rate loss of the system is smaller than 0.06dB and 0.03bits/s/Hz when the number of quantization bits are equal to 4 and 3, respectively. The achievable rates of the system in the no performance loss, performance loss, and approximate performance loss cases increase with the number of the phase shift elements of IRS. When the number of quantization bits is larger than 1,  regardless of the value of the number of quantization bits, the difference between the performance losses and approximate performance losses of the SNR and achievable rate is trivial.

\end{enumerate}

The remainder of this work is organized below. We show the system model of double IRS-assisted AF relay network in Section \ref{s2}.
In Sections \ref{s3}, the performance loss expressions are derived for the Rician channels. The numerical simulation results and conclusions are presented in Section \ref{s4} and Section \ref{s5}, respectively.

{\bf Notations:} throughout this paper, lower case, boldface lower case,  and boldface upper case letters refer to scalars, vectors, and matrices, respectively. The notations $\mathbb{E}(\cdot)$, $(\cdot)^T$, $(\cdot)^H$, and $|\cdot|_m$ mean the mathematic expectation, transpose, conjugate transpose, and $m$-th element absolute value, respectively. The sign $\mathbb{C}^{N\times N}$ denotes the matrix space of $N\times N$.

\section{system model}\label{s2}
As indicated in Fig. 1, a double IRS-assisted AF relay station (RS) wireless network is considered, where AF relay operates in half-duplex mode. The Source (S) delivers confidential message to the Destination (D) with the aid of IRS-1, IRS-2, and RS. It is assumed that there is no direct confidential message (CM) transmission between S and D, between S and IRS-2, and between IRS-1 and D because of the long distance. Herein, S, D, and RS are equipped with single antenna. The IRS-1 and IRS-2 are equipped with $N$ and $M$ passive reflecting elements, respectively, and they reflect signal only one time slot. $\textbf{h}_{si} \in \mathbb{C}^{N\times 1}$, $\textbf{h}_{ir}^H \in \mathbb{C}^{1 \times N}$, $h_{sr}^H \in \mathbb{C}^{1\times 1}$, $\textbf{h}_{ri} \in \mathbb{C}^{M\times 1}$, $\textbf{h}_{id}^H \in \mathbb{C}^{1 \times M}$, $h_{rd}^H \in \mathbb{C}^{1 \times 1}$ are the S$\rightarrow$IRS-1, IRS-1$\rightarrow$RS, S$\rightarrow$RS, RS$\rightarrow$IRS-2, IRS-2$\rightarrow$D, and RS$\rightarrow$D channels, respectively.

\begin{figure}[htpt]
\centering
\includegraphics[width=0.45\textwidth]{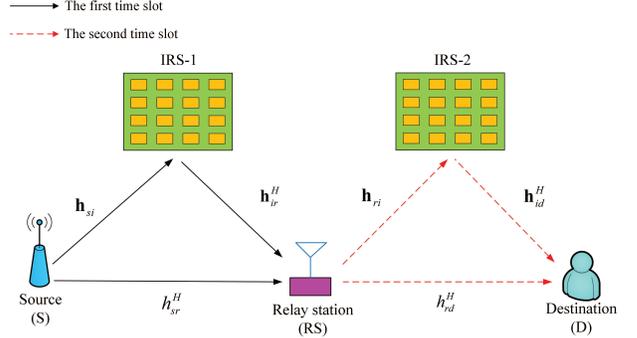}\\
\caption{Block diagram for IRS-aided amplify-and-forward relay network.}\label{System-model.eps}
\end{figure}


In the first time slot, S sends CM to RS with the help of IRS-1, and the received signal of RS is
\begin{equation}\label{y_r1}
{y_r}=\sqrt{{g_{sr}}{P_s}}{{h}_{sr}^H}{x_s}+\sqrt{{g_{sir}}{P_s}}{\textbf{h}_{ir}^H}{\boldsymbol{\Theta}_1}{\textbf{h}_{si}}{x_s}+{n_r},
\end{equation}
where $x_s$ and $P_s$ stand for the transmit signal and power of S, respectively, $g_{sir}=g_{si}g_{ir}$ means the equivalent path loss coefficient of S$\rightarrow$IRS-1 and IRS-1$\rightarrow$RS channels, and $g_{sr}$ means the path loss coefficient of S$\rightarrow$RS channel. $\boldsymbol{\Theta}_1=\text{diag}(e^{j\phi_{1}},\cdots, e^{j\phi_{n}}, \cdots, e^{j\phi_{N}})$ stands for the diagonal reflection coefficient matrix of IRS-1, $\phi_{n}\in (0, 2\pi]$ denotes the phase shift of the $n$-th reflection element. $n_r$ stands for the additive white Gaussian noise (AWGN) of RS with the distribution $n_r\sim\mathcal {C}\mathcal {N}(0, \sigma^2_{r})$.

\section{Performance loss analysis}\label{s3}
Assuming that all channels are the Rayleigh channels in this section. The IRS with discrete phase shifters might introduce phase quantization errors and degrade system performance. In what follows, we will derive the closed-form expressions for SNR and AR with respect to the number of quantized bits, respectively, and analyze the effect of IRS with discrete phase shifters on the system performance in the first and second time slots.

\subsection{In the first time slot}
Assuming that all Rayleigh channels obey the Rayleigh distribution, and the corresponding probability density function is given by
\begin{equation}\label{f}
  f_{\alpha}(x)=\left\{
             \begin{array}{ll}
             \frac{x}{{\alpha}^2}e^{- \frac{x^2}{2{\alpha}^2}}, & \hbox{$x\in[0,+\infty)$,}\\
             0, & \hbox{$x\in(-\infty, 0)$},
             \end{array}
           \right.
\end{equation}
where $\alpha$ stands for the Rayleigh distribution parameter and $\alpha > 0$.

The received signal (\ref{y_r1}) can be recast as
\begin{align}\label{y_rlos}
&{y_r}=\sqrt{g_{sr}P_s}{{h}_{sr}^H}{x_s}+\sqrt{g_{sir}P_s}{\textbf{h}_{ir}^H}{\boldsymbol{\Theta}_1}{\textbf{h}_{si}}{x_s}+{n_r}\nonumber\\
&=\sqrt{g_{sr}P_s}\left|{h_{sr}}\right|{e^{-j\varphi_{sr}}}{x_s}+\sqrt{g_{sir}}\sum\limits_{n=1}^N\left|h_{ir}(n)\right|\left|h_{si}(n)\right|\cdot\nonumber\\
&~~~e^{j(-2\pi\Psi_{\theta_{ir}}(n)+\phi_n+2\pi\Psi_{\theta_{si}}(n))}x_s+n_r\nonumber\\
&=\Big(\sqrt{\frac{g_{sir}}{g_{sr}}}\sum\limits_{n=1}^N{e^{j(-2\pi\Psi_{\theta_{ir}}(n)+\phi_n+2\pi\Psi_{\theta_{si}}(n)+\varphi_{sr})}}\cdot\nonumber\\
&~~~ \left|h_{ir}(n)\right|\left|h_{si}(n)\right|+|{h_{sr}}|\Big)\sqrt{g_{sr}P_s}{e^{-j\varphi_{sr}}}x_s+n_r,
\end{align}
where $\varphi_{sr}$ represents the phase of $h_{sr}$. For convenience of derivation, we assume $\varphi_{sr}=0$ and
\begin{equation}
\phi_n=-\varphi_{sr}+2\pi\Psi_{\theta_{ir}}(n)-2\pi\Psi_{\theta_{si}}(n).
\end{equation}
At this point, based on the law of weak large numbers in \cite{Wasserman2004All} and Rayleigh distribution, and due to the fact that $\left|h_{ir}(n)\right|$ and $\left|h_{si}(n)\right|$ are independent identically distributed Rayleigh distributions, whose parameters respectively are $\alpha_{ir}$ and $\alpha_{si}$, then, (\ref{y_rlos}) can be simplified to
\begin{align}\label{r5}
y_r&=\sqrt{P_s}(\sqrt{g_{sr}}|h_{sr}|+\sqrt{g_{sir}}\sum\limits_{n=1}^N|h_{ir}(n)||h_{si}(n)|)x_s+n_r\nonumber\\
&=\sqrt{P_s}(\sqrt{g_{sr}}\mathbb{E}(|h_{sr}|)+\sqrt{g_{sir}}N\mathbb{E}(|h_{ir}(n)||h_{si}(n)|)x_s\nonumber\\
&~~~+n_r\nonumber\\
&=\sqrt{P_s}\Big(\sqrt{g_{sr}}\int_{0}^{+\infty}|h_{sr}|f_{\alpha_{sr}}(|h_{sr}|)d(|h_{sr}|)+\sqrt{g_{sir}}N\cdot\nonumber\\
&~~~\int_{0}^{+\infty}|h_{ir}(n)|f_{\alpha_{ir}}(|h_{ir}(n)|)d(|h_{ir}(n)|)\int_{0}^{+\infty}|h_{si}(n)|\cdot\nonumber\\
&~~~f_{\alpha_{si}}(|h_{si}(n)|)d(|h_{si}(n)|\Big)x_s+n_r\nonumber\\
&=\sqrt{P_s}\Big(\sqrt{\frac{\pi}{2}g_{sr}}\alpha_{sr}+\sqrt{g_{sir}}N\frac{\pi}{2}\alpha_{ir}\alpha_{si}\Big)x_s+n_r,
\end{align}
where $\alpha_{sr}$ is the Rayleigh distribution parameter of S-to-RS channel.

It is assumed that the $k$-bit phase quantizer is adopted by the IRS with discrete phase shifter, the phase feasible set of each reflection element of IRS is
\begin{align}\label{Omega}
\Omega=\left\{ \frac{1}{2^k}\pi, \frac{3}{2^k}\pi, \cdots, \frac{2^{k+1}-1}{2^k}\pi \right\}.
\end{align}
Assuming that the actual discrete phase $\overline{\phi_n}$ of the $n$-th element at IRS is chosen from the phase feasible set $\Omega$ in (\ref{Omega}), denoted as
\vspace{-0.1cm}
\begin{align}
\overline{\phi_n}=\mathop{\arg\min}\limits_{{\overline{\phi_n}\in \Omega}} \|\overline{\phi_n}-\phi_n\|_2,
\end{align}
where $\phi_n$ is the ideal continuous phase, In general, $\overline{\phi_n} \neq \phi_n$, it implies the phase mismatching, which may result in performance loss at RS. Defining the quantization error of the $n$-th phase at IRS as
\begin{equation}\label{Delta}
\Delta \phi_n=\overline{\phi_n}-\phi_n.
\end{equation}
Assuming that $\Delta \phi_n$ obeys the uniform distribution, its probability density function is
\begin{equation}\label{f1}
  f(x)=\left\{
             \begin{array}{ll}
             \frac{1}{2\Delta x}, & \hbox{$x\in[-\Delta x,\Delta x]$,}\\
             0, & \hbox{otherwise},
             \end{array}
           \right.
\end{equation}
where $\Delta x=\frac{1}{2^k}\pi$.

When employing the IRS with discrete phase shifters, there is a phase mismatch, which leads to phase quantization error and performance loss. When there is a phase quantization error, the received signal (\ref{y_rlos}) can be recast as
\begin{align}\label{hyrr2}
\hat{y}_r&=\sqrt{g_{sr}P_s}{{h}_{sr}^H}{x_s}+\sqrt{g_{sir}P_s}{\textbf{h}_{ir}^H}{\boldsymbol{\Theta}_1}{\textbf{h}_{si}}{x_s}+{n_r}\nonumber\\
&=\sqrt{g_{sr}P_s}\left|{h_{sr}}\right|{e^{-j\varphi_{sr}}}{x_s}+\sqrt{g_{sir}}\sum\limits_{n=1}^N\left|h_{ir}(n)\right|\left|h_{si}(n)\right|\cdot\nonumber\\
&~~~e^{j\overbrace{(-2\pi\Psi_{\theta_{ir}}(n)+\phi_n+2\pi\Psi_{\theta_{si}}(n))}^{\Delta \phi_n}}x_s+n_r\nonumber\\
&=\sqrt{P_s}(\sqrt{g_{sr}}\left|h_{sr}\right|+\sqrt{g_{sir}}\sum\limits_{n=1}^N\left|h_{ir}(n)\right|\left|h_{si}(n)\right|{e^{j\Delta \phi_n}})x_s\nonumber\\
&~~~+n_r.
\end{align}
Due to the fact that $\left|h_{ir}(n)\right|$, $\left|h_{si}(n)\right|$, and $e^{j\Delta \phi_n}$ are independent of each other, by using the same method as in (\ref{r5}) and Euler formula, i.e.,
\begin{align} 
e^{j\Delta \phi_n}=\text{cos}(\Delta \phi_n)+j\text{sin}(\Delta \phi_n),
\end{align}
(\ref{hyrr2}) can be rewritten as
\begin{align}\label{y3}
\hat{y}_r&=\sqrt{P_s}(\sqrt{g_{sr}}\int_{0}^{+\infty}|h_{sr}||f_{\alpha_{sr}}(|h_{sr}|)d(|h_{sr}|)+\sqrt{g_{sir}}N\cdot\nonumber\\
&~~~\int_{0}^{+\infty}|h_{ir}(n)||f_{\alpha_{ir}}(|h_{ir}(n)|)d(|h_{ir}(n)|)\int_{0}^{+\infty}|h_{si}(n)|\cdot\nonumber\\
&~~~f_{\alpha_{si}}(|h_{si}(n)|)d(|h_{si}(n)|\int_{-\Delta x}^{\Delta x}(\text{cos}(\Delta \phi_n)d(\Delta \phi_n)+\nonumber\\
&~~~j\text{sin}(\Delta \phi_n)d(\Delta \phi_n))x_s+n_r\nonumber\\
&=\sqrt{P_s}\Big(\sqrt{\frac{\pi}{2}g_{sr}}\alpha_{sr}+\sqrt{g_{sir}}N\text{sinc}\left(\frac{\pi}{2^{k_1}}\right)\frac{\pi}{2}\alpha_{ir}\alpha_{si}\Big)x_s\nonumber\\
&~~~+n_r,
\end{align}
where $k_1$ stands for the number of quantization bits of IRS-1 and is a finite positive integer. 

In the following, to simplify (\ref{y3}) and given that $\Delta \phi_n \rightarrow 0$ when the number $k_1$ of quantized bits becomes large, by the Taylor series expansion, the approximation can be obtained as follows
\begin{align}\label{cos1}
\text{cos}(\Delta \phi_n) \approx 1-\frac{\Delta \phi_n^2}{2}.
\end{align}
Substituting (\ref{cos1}) into (\ref{y3}), the received signal (\ref{y_rlos}) with approximate phase quantization error can be obtained as
\begin{align}\label{APL1}
\widetilde{y}_r&=\sqrt{P_s}\Big(\sqrt{\frac{\pi}{2}g_{sr}}\alpha_{sr}+\sqrt{g_{sir}}N\Big(1-\frac{1}{6}{\left(\frac{\pi}{2^{k_1}}\right)}^2\Big)\cdot\nonumber\\
&~~~\frac{\pi}{2}\alpha_{ir}\alpha_{si}\Big)x_s+n_r.
\end{align}

\subsection{In the second time slot}
Based on (\ref{y_r1}), assuming that RS receives the signal and then successfully amplifies and forwards it, the received signal at D can be written as
\begin{equation}\label{y_d11}
{y_d}=\sqrt{g_{rd}P_r}{{h}_{rd}^H}{x_r}+\sqrt{g_{rid}P_r}{\textbf{h}_{id}^H}{\boldsymbol{\Theta}_2}{\textbf{h}_{ri}}{x_r}+{n_d},
\end{equation}
where $x_r$ and $P_r$ stand for the transmit signal and power of RS, respectively, $g_{rid}=g_{ri}g_{id}$ indicates the equivalent path loss coefficient of RS$\rightarrow$IRS-2 and IRS-2$\rightarrow$D channels, and $g_{rd}$ means the path loss coefficient of RS$\rightarrow$D channel. $\boldsymbol{\Theta}_2=\text{diag}(e^{j\phi_{1}}, \cdots, e^{j\phi_{m}}, \cdots, e^{j\phi_{M}})$ means the reflection coefficient matrix of IRS-2, where $\phi_{m}\in (0, 2\pi]$ stands for the phase shift of reflection element $m$. $n_d$ denotes the AWGN at D with the distribution $n_d\sim\mathcal {C}\mathcal {N}(0, \sigma^2_{d})$.
The expression of $x_r$ is given by
\begin{align}\label{x_r}
x_r &=\beta{y_r}\nonumber\\
&=\beta (\sqrt{g_{sr}P_s}{{h}_{sr}^H}{x_s}+\sqrt{g_{sir}P_s}{\textbf{h}_{ir}^H}{\boldsymbol{\Theta}_1}{\textbf{h}_{si}}{x_s}+{n_r})\nonumber\\
&=\beta \sqrt{P_s}\left(\sqrt{g_{sr}}h_{sr}^H+\sqrt{g_{sir}}{\textbf{h}_{ir}^H}{\boldsymbol{\Theta}_1}{\textbf{h}_{si}}\right)x_s+\beta n_r.
\end{align}
The amplification factor of AF relay is
\begin{align}\label{beta1}
\beta =\frac{\sqrt{P_r}}{\sqrt{{P_s}{\left|\sqrt{g_{sr}}h_{sr}^H+{\sqrt{g_{sir}}\textbf{h}_{ir}^H}{\boldsymbol{\Theta}_1}{\textbf{h}_{si}}\right|}^2+{\sigma}_r^2}}.
\end{align}

When there is no phase quantization error, according to (\ref{r5}), the amplification factor of AF relay (\ref{beta1}) can be recast as
\begin{align}
\beta =\frac{\sqrt{P_r}}{\sqrt{{P_s}{\Big(\sqrt{\frac{\pi}{2}g_{sr}}\alpha_{sr}
+\sqrt{g_{sir}}N\frac{\pi}{2}\alpha_{ir}\alpha_{si}\Big)}^2+{\sigma}_r^2}}.
\end{align}
In the same manner, based on (\ref{y3}) and (\ref{APL1}), the AF relay amplification factors with quantization error and approximate quantization error are
\begin{align}
\hat\beta =\frac{\sqrt{P_r}}{\sqrt{{P_s}{\Big(\sqrt{\frac{\pi}{2}g_{sr}}\alpha_{sr}
+\sqrt{g_{sir}}N\text{sinc}\left(\frac{\pi}{2^{k_1}}\right)\frac{\pi}{2}\alpha_{ir}\alpha_{si}\Big)}^2+{\sigma}_r^2}}
\end{align}
and
\begin{align}
&\widetilde\beta =\nonumber\\
&\frac{\sqrt{P_r}}{\sqrt{{P_s}{\left(\sqrt{\frac{\pi}{2}g_{sr}}\alpha_{sr}
+\sqrt{g_{sir}}N\left(1-\frac{1}{6}{\left(\frac{\pi}{2^{k_1}}\right)}^2\right)\frac{\pi}{2}\alpha_{ir}\alpha_{si}\right)}^2+{\sigma}_r^2}},
\end{align}
respectively. 

When there is no phase quantization error, (\ref{y_d11}) can be rewritten as
\begin{align}\label{ydrl3}
y_d&=\beta\sqrt{{P_r}{P_s}}\Bigg(\sqrt{g_{sr}g_{rd}}\frac{\pi}{2}\alpha_{rd}\alpha_{sr}+\sqrt{g_{sir}g_{rd}}N{\Big(\frac{\pi}{2}\Big)}^{\frac{3}{2}}\cdot\nonumber\\
&~~~\alpha_{rd}\alpha_{ir}\alpha_{si}+\sqrt{g_{sr}g_{rid}}M{\Big(\frac{\pi}{2}\Big)}^{\frac{3}{2}}\alpha_{id}\alpha_{ri}\alpha_{sr}+\nonumber\\
&~~~\sqrt{g_{sir}g_{rid}}MN\frac{{\pi}^2}{4}\alpha_{id}\alpha_{ri}\alpha_{ir}\alpha_{si}\Bigg)x_s+\nonumber\\
&~~~\beta \sqrt{P_r}\Big(\sqrt{\frac{\pi}{2}g_{rd}}\alpha_{rd}+\sqrt{g_{rid}}M\frac{\pi}{2}\alpha_{id}\alpha_{ri}\Big)n_r+n_d.
\end{align}
In the cases with phase quantization error and approximate phase quantization error, the received signal at D can be recast as
\begin{align}\label{hydrl3}
\hat{y}_d=&\hat{\beta}\sqrt{{P_r}{P_s}}\Bigg(\sqrt{g_{sr}g_{rd}}\frac{\pi}{2}\alpha_{rd}\alpha_{sr}
+\sqrt{g_{sir}g_{rd}}N{\Big(\frac{\pi}{2}\Big)}^{\frac{3}{2}}\nonumber\\
&\alpha_{rd}\alpha_{ir}\alpha_{si}\text{sinc}\left(\frac{\pi}{2^{k_1}}\right)+\sqrt{g_{sr}g_{rid}}M{\Big(\frac{\pi}{2}\Big)}^{\frac{3}{2}}\alpha_{id}\alpha_{ri}\cdot\nonumber\\
&\alpha_{sr}\text{sinc}\left(\frac{\pi}{2^{k_2}}\right)+\sqrt{g_{sir}g_{rid}}MN\frac{{\pi}^2}{4}\alpha_{id}\alpha_{ri}\alpha_{ir}\alpha_{si}\cdot\nonumber\\
&\text{sinc}\left(\frac{\pi}{2^{k_1}}\right)\text{sinc}\left(\frac{\pi}{2^{k_2}}\right)\Bigg)x_s+\hat{\beta} \sqrt{P_r}\Bigg(\sqrt{\frac{\pi}{2}g_{rd}}\alpha_{rd}+\nonumber\\
&\sqrt{g_{rid}}M\frac{\pi}{2}\alpha_{id}\alpha_{ri}\text{sinc}\left(\frac{\pi}{2^{k_2}}\right)\Bigg)n_r+n_d
\end{align}
and
\begin{align}\label{wydrl3}
\widetilde{y}_d=&\widetilde{\beta}\sqrt{{P_r}{P_s}}\Bigg(\sqrt{g_{sr}g_{rd}}\frac{\pi}{2}\alpha_{rd}\alpha_{sr}+\sqrt{g_{sir}g_{rd}}N{\Big(\frac{\pi}{2}\Big)}^{\frac{3}{2}}\cdot\nonumber\\
&\alpha_{rd}\alpha_{ir}\alpha_{si}\Big(1-\frac{1}{6}{\Big(\frac{\pi}{2^{k_1}}\Big)}^2\Big)+\sqrt{g_{sr}g_{rid}}M{\Big(\frac{\pi}{2}\Big)}^{\frac{3}{2}}\cdot\nonumber\\
&\alpha_{id}\alpha_{ri}\alpha_{sr}\Big(1-\frac{1}{6}{\Big(\frac{\pi}{2^{k_2}}\Big)}^2\Big)+\sqrt{g_{sir}g_{rid}}MN\frac{{\pi}^2}{4}\cdot\nonumber\\
&\alpha_{id}\alpha_{ri}\alpha_{ir}\alpha_{si}\Big(1-\frac{1}{6}{\Big(\frac{\pi}{2^{k_1}}\Big)}^2\Big)
\Big(1-\frac{1}{6}{\Big(\frac{\pi}{2^{k_2}}\Big)}^2\Big)\Bigg)x_s\nonumber\\
&+\widetilde{\beta} \sqrt{P_r}\Bigg(\sqrt{\frac{\pi}{2}g_{rd}}\alpha_{rd}+\sqrt{g_{rid}}M\frac{\pi}{2}\alpha_{id}\alpha_{ri}\cdot\nonumber\\
&\Big(1-\frac{1}{6}{\Big(\frac{\pi}{2^{k_2}}\Big)}^2\Big)\Bigg)n_r+n_d,
\end{align}
respectively, where $k_2$ stands for the number of quantization bits of IRS-2.

Let us define
\begin{align}\label{v1}
v_1=\sqrt{g_{sir}g_{rd}}N{\Big(\frac{\pi}{2}\Big)}^{\frac{3}{2}}\alpha_{rd}\alpha_{ir}\alpha_{si},
\end{align}
\begin{align}
v_2=\sqrt{g_{sr}g_{rid}}M{\Big(\frac{\pi}{2}\Big)}^{\frac{3}{2}}\alpha_{id}\alpha_{ri}\alpha_{sr},
\end{align}
\begin{align}
v_3=\sqrt{g_{sir}g_{rid}}MN\frac{{\pi}^2}{4}\alpha_{id}\alpha_{ri}\alpha_{ir}\alpha_{si},
\end{align}
\begin{align}\label{v4}
v_4={\beta}^2{P_r}{\Bigg(\sqrt{\frac{\pi}{2}g_{rd}}\alpha_{rd}+\sqrt{g_{rid}}M\frac{\pi}{2}\alpha_{id}\alpha_{ri}\Bigg)}^2\sigma_r^2+\sigma_d^2.
\end{align}
In accordance with (\ref{ydrl3}) and (\ref{v1})-(\ref{v4}), the expression of SNR without performance loss at D is
\begin{align}
\text{SNR}_d=\frac{{\beta}^2{P_r}P_s{\Big(\sqrt{g_{sr}g_{rd}}\frac{\pi}{2}\alpha_{rd}\alpha_{sr}+v_1+v_2+v_3\Big)}^2}{v_4}.
\end{align}
Let
\begin{align}
&u_1=v_1\text{sinc}\left(\frac{\pi}{2^{k_1}}\right),~
u_2=v_2\text{sinc}\left(\frac{\pi}{2^{k_2}}\right),\\
&u_3=v_3\text{sinc}\left(\frac{\pi}{2^{k_1}}\right)\text{sinc}\left(\frac{\pi}{2^{k_2}}\right),
\end{align}
\begin{align}
u_4=&\hat{\beta}^2{P_r}{\left(\sqrt{\frac{\pi}{2}g_{rd}}\alpha_{rd}+
\sqrt{g_{rid}}M\frac{\pi}{2}\alpha_{id}\alpha_{ri}\text{sinc}\left(\frac{\pi}{2^{k_2}}\right)\right)}^2\sigma_r^2\nonumber\\
&+\sigma_d^2,
\end{align}
then, the SNR expression with performance loss at D is
\begin{align}
\widehat{\text{SNR}}_d=\frac{\hat{\beta}^2{P_r}P_s{\Big(\sqrt{g_{sr}g_{rd}}\frac{\pi}{2}\alpha_{rd}\alpha_{sr}+u_1+u_2+u_3\Big)}^2}{u_4}.
\end{align}
Defining
\begin{align}\label{q1}
&q_1=v_1\Big(1-\frac{1}{6}{\Big(\frac{\pi}{2^{k_1}}\Big)}^2\Big),~
q_2=v_2\Big(1-\frac{1}{6}{\Big(\frac{\pi}{2^{k_2}}\Big)}^2\Big),\\
&q_3=v_3\Big(1-\frac{1}{6}{\Big(\frac{\pi}{2^{k_1}}\Big)}^2\Big)\Big(1-\frac{1}{6}{\Big(\frac{\pi}{2^{k_2}}\Big)}^2\Big),\\
&q_4=\widetilde{\beta}^2{P_r}\Bigg(\sqrt{\frac{\pi}{2}g_{rd}}\alpha_{rd}+\sqrt{g_{rid}}M\frac{\pi}{2}\alpha_{id}\alpha_{ri}\cdot\nonumber\\
&~~~~~~\left(1-\frac{1}{6}{\left(\frac{\pi}{2^{k_2}}\right)}^2\right)\Bigg)^2\sigma_r^2+\sigma_d^2.\label{q4}
\end{align}
According to (\ref{wydrl3}) and (\ref{q1})-(\ref{q4}), the SNR expression with approximate performance loss at D can be obtained by
\begin{align}
\widetilde{\text{SNR}}_d=\frac{\widetilde{\beta}^2{P_r}P_s{\Big(\sqrt{g_{sr}g_{rd}}\frac{\pi}{2}\alpha_{rd}\alpha_{sr}+q_1+q_2+q_3\Big)}^2}{q_4}.
\end{align}
The performance loss and approximate performance loss of SNR at D are given by
\begin{align}
\hat{L}_d&=\frac{\text{SNR}_d}{\widehat{\text{SNR}}_d}\nonumber\\
&=\frac{u_4{\beta}^2{\Big(\sqrt{g_{sr}g_{rd}}\frac{\pi}{2}\alpha_{rd}\alpha_{sr}+v_1+v_2+v_3\Big)}^2}
{v_4\hat{\beta}^2{\Big(\sqrt{g_{sr}g_{rd}}\frac{\pi}{2}\alpha_{rd}\alpha_{sr}+u_1+u_2+u_3\Big)}^2}
\end{align}
and
\begin{align}
\widetilde{L}_d&=\frac{\text{SNR}_d}{\widetilde{\text{SNR}}_d}\nonumber\\
&=\frac{q_4{\beta}^2{\Big(\sqrt{g_{sr}g_{rd}}\frac{\pi}{2}\alpha_{rd}\alpha_{sr}+v_1+v_2+v_3\Big)}^2}
{v_4\widetilde{\beta}^2{\Big(\sqrt{g_{sr}g_{rd}}\frac{\pi}{2}\alpha_{rd}\alpha_{sr}+q_1+q_2+q_3\Big)}^2},
\end{align}
respectively.

Correspondingly, according to (\ref{ydrl3}), (\ref{hydrl3}), and (\ref{wydrl3}), the AR expressions with no performance loss, with performance loss, and with approximate performance loss at D are given by
\begin{align}
&{R}_d=\nonumber\\
&{\log}_2\Bigg(1+\frac{{\beta}^2P_rP_s{\left(\sqrt{g_{sr}g_{rd}}\frac{\pi}{2}\alpha_{rd}\alpha_{sr}+v_1+v_2+v_3\right)}^2}
{v_4}\Bigg),
\end{align}
\begin{align}
&\hat{R}_d=\nonumber\\
&{\log}_2\Bigg(1+\frac{{\hat{\beta}}^2P_rP_s{\left(\sqrt{g_{sr}g_{rd}}\frac{\pi}{2}\alpha_{rd}\alpha_{sr}+u_1+u_2+u_3\right)}^2}
{u_4}\Bigg),
\end{align}
and
\begin{align}
&\widetilde{R}_d=\nonumber\\
&{\log}_2\Bigg(1+\frac{{\widetilde{\beta}}^2P_rP_s{\left(\sqrt{g_{sr}g_{rd}}\frac{\pi}{2}\alpha_{rd}\alpha_{sr}+q_1+q_2+q_3\right)}^2}
{q_4}\Bigg),
\end{align}
respectively.

\section{SIMULATION RESULTS}\label{s4}
In this section, we present the simulation results to analyze the impact of phase mismatch induced by IRS with discrete phase shifter in terms of both SNR and AR. Assume that the path loss model with distance $\overline{d}_0$ is $g(\overline{d}_0)=PL_0-10\gamma {\log}_{10}\overline{d}_0/d_0$, where $PL_0=-30$dBm indicates the path loss at the reference distance $d_0=1$m. $\gamma$ stands for the path loss index. In the Rayleigh channels, the path loss exponents of S$\rightarrow$IRS-1, IRS-1$\rightarrow$-RS, S$\rightarrow$RS, RS$\rightarrow$IRS-2, IRS-2$\rightarrow$D and RS$\rightarrow$D channels are 2.6, 2.6, 3.5, 2.6, 2.6, 3.5, respectively. The system parameters are listed as follows: $d=\lambda /2$, $\theta_{sr}=\theta_{rd}=\pi/2$, $\theta_{si}=\theta_{ri}=\pi/4$, $d_{si}=d_{ri}=50$m, $d_{sr}=d_{rd}=150$m, $P_s=30$dBm, and $P_r=35$dBm. The Rayleigh distribution parameters for all Rayleigh channels are set as 0.5.

\begin{figure*}
 \setlength{\abovecaptionskip}{-5pt}
 \setlength{\belowcaptionskip}{-10pt}
 \centering
 \begin{minipage}[t]{0.33\linewidth}
  \centering
  \includegraphics[width=2.56in]{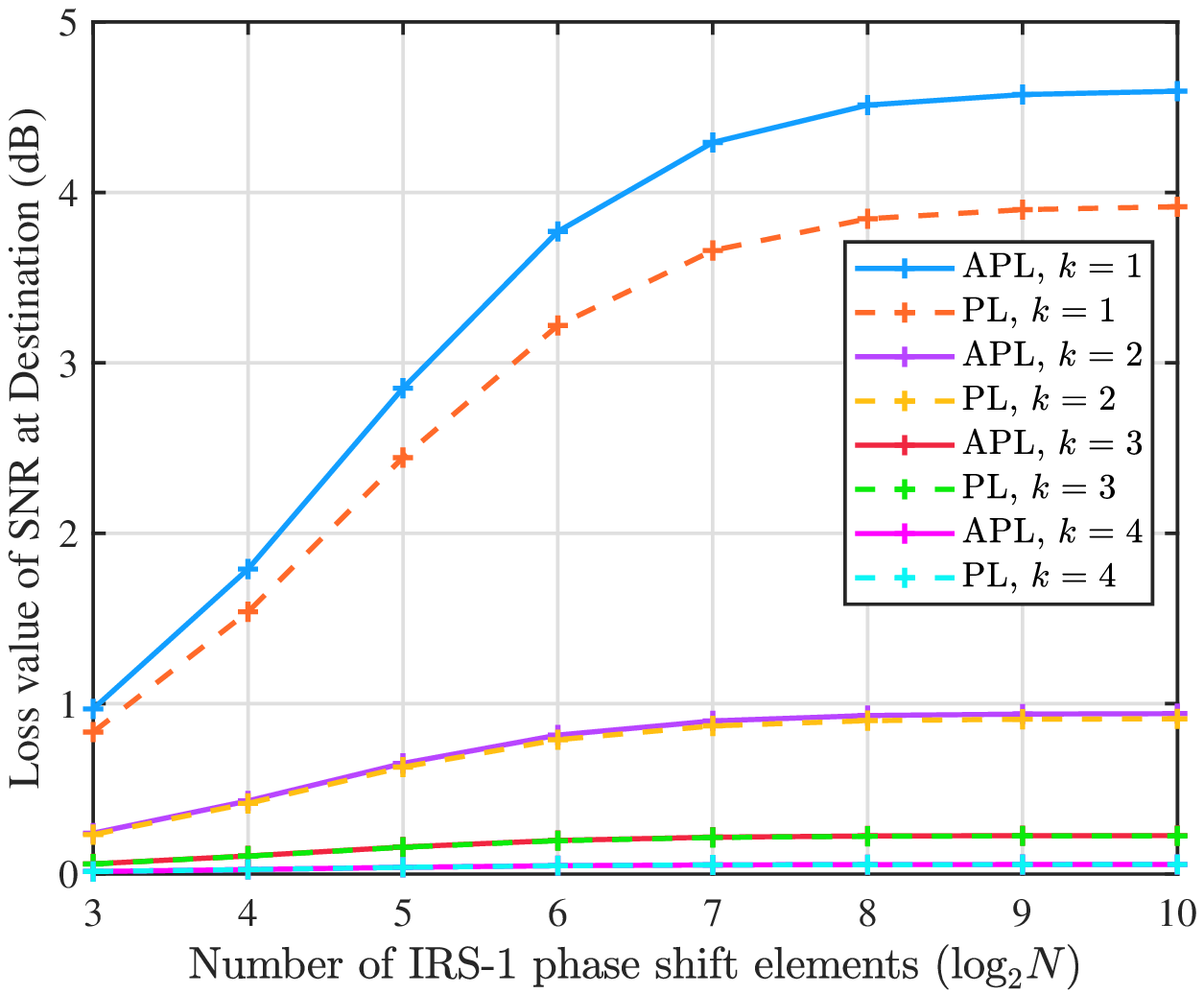}
  \caption{SNR performance loss versus $N$.}\label{SNR}
 \end{minipage}%
 \begin{minipage}[t]{0.33\linewidth}
  \centering
  \includegraphics[width=2.56in]{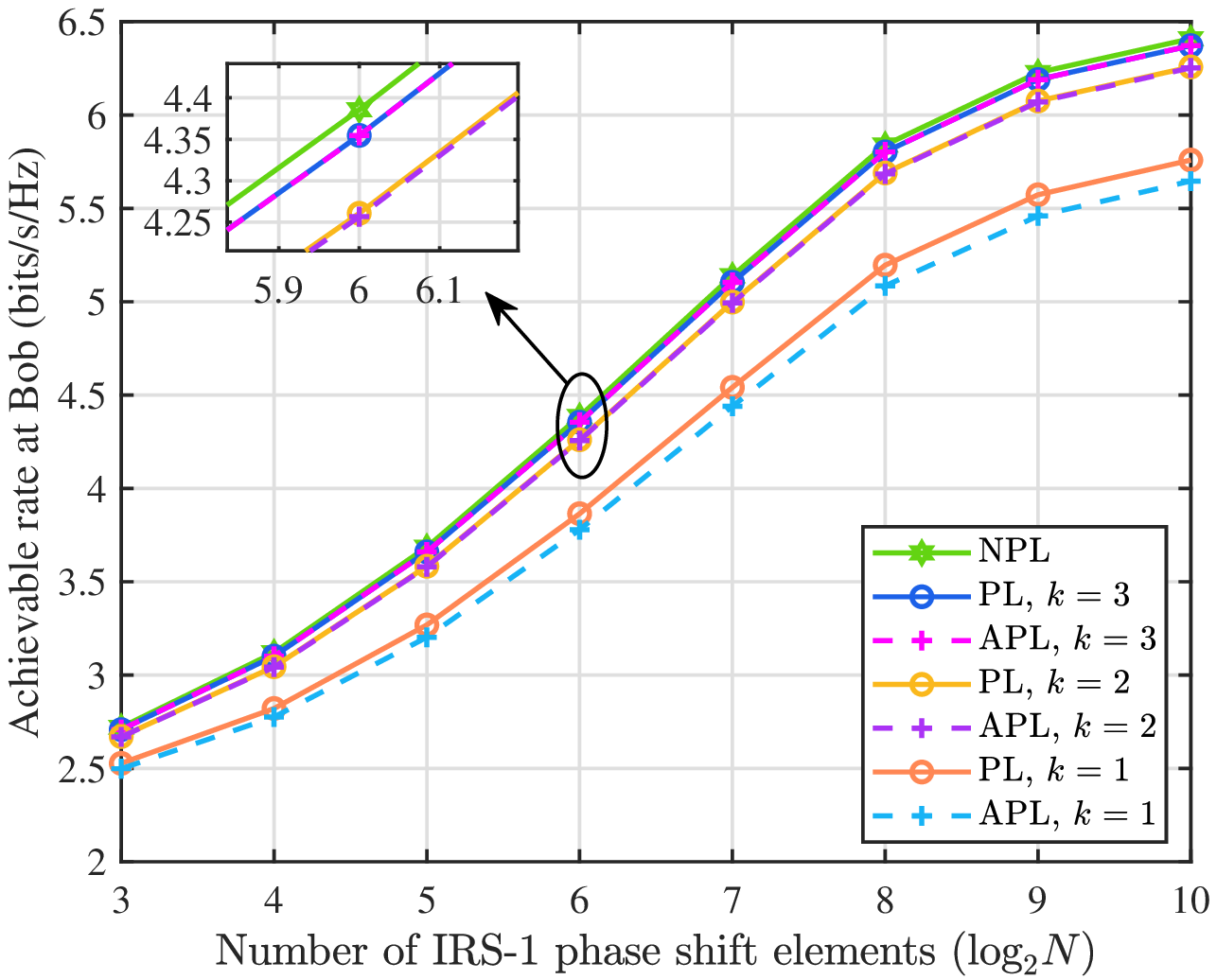}
  \caption{Achievable rate versus $N$.}\label{AR_L}
 \end{minipage}
 \begin{minipage}[t]{0.33\linewidth}
  \centering
  \includegraphics[width=2.56in]{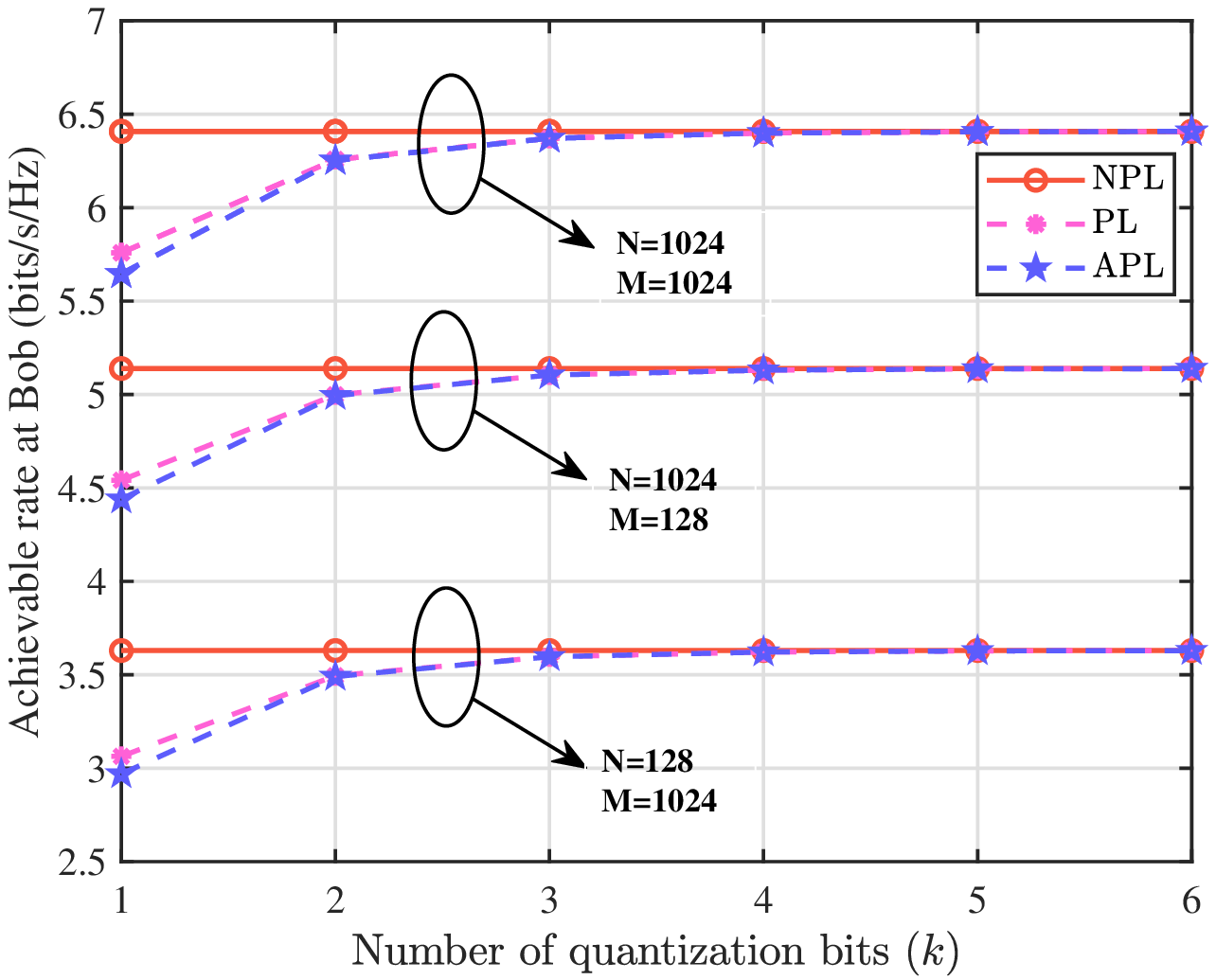}
  \caption{Achievable rate versus $k$.}\label{AR_k}
 \end{minipage}
\end{figure*}

Fig. \ref{SNR} shows the SNR performance loss versus the number $N$ of IRS-1 phase shift elements at Destination, where the number of IRS-2 phase shift elements $M=N$. It can be seen from the figure that regardless of the scenarios of performance loss (PL) or approximate performance loss (APL), the SNR performance loss decreases with $k$, and increase with $N$. The difference between PL and APL increases with the increases of $N$ when $k$ is equal to 1, and is trivial when $k$ is larger than or equal to 2. When $k=4$, the SNR performance loss of the system is lower than 0.06dB. Moreover, as $N$ tends to large scale, the increase in the system's SNR PL and APL gradually slows down.

Fig. \ref{AR_L} depicts the curve of AR versus the number $N$ of IRS-1 phase shift elements at Destination, where the number of IRS-2 phase shift elements $M=N$. As can be found from the figure that the ARs of the system in the no performance loss (NPL), PL, and APL cases increase gradually with $N$. The difference of ARs between the PL case and the APL case increases gradually with $N$ when $k=1$, and is negligible when $k=2$. The achievable rates in both the PL and APL cases gradually approach the ones in the NPL case as $k$ increases. For example, when $N=1024$, the AR PL is about 0.15bits/s/Hz at $k=2$, and only 0.03bits/s/Hz at $k=3$.

Fig. \ref{AR_k} illustrates the AR versus the number $k$ of quantization bits form 1 to 6. As shown in this figure, the performance loss of AR at Destination decreases as $k$ increases, while increases with the number $N$ of IRS-1 phase shift elements and number $M$ of IRS-2 phase shift elements increase. Compared to the scenario of NPL, when $k=3$, the AR performance loss is less than 0.04bits/s/Hz with both PL and APL. When $k=1$, the difference in ARs among the NPL, PL and APL cases increases gradually with the increase of $N$ and $M$. Whether in NPL, PL, or APL case, the AR at $N = 1024, M = 128$ is higher than that at $N = 128, M = 1024$. This reveals that increasing the value of $N$ has a more significant enhancement on the AR performance than that the value of $M$.

\section{Conclusions}\label{s5}
In this paper, we have made an analysis of the the performance loss of the double IRS-assisted AF relay network due to the phase quantization error of IRS with discrete phase shifters in the Rayleigh channels. According to the law of large numbers, Euler formula, and Rayleigh distribution, the expressions of the SNR performance loss and achievable rate were derived. In accordance with the Taylor series expansion, their approximate performance loss expressions were also derived. Simulation results showed that both the SNR and achievable rate performance losses and approximate performance losses decrease gradually with increasing number of quantization bits, and increase gradually with increasing numbers of IRS phase shift elements. The SNR performance loss and achievable rate performance loss of the system are trivial when the number of quantization bits is equal to 4 and 3, respectively. Therefore, 4-bit quantizer is sufficient to enable a negligible performance loss.

\ifCLASSOPTIONcaptionsoff
  \newpage
\fi

\bibliographystyle{IEEEtran}
\bibliography{IEEEfull,reference}
\end{document}